\documentclass[12pt,3p,times]{elsarticle}

\usepackage{graphicx}

\usepackage[hyphenbreaks]{breakurl}
\usepackage[hyphens]{url}

\usepackage{smartdiagram}
\smartdiagramset{set color list={
		gray!30,
		gray!30,
		gray!30,
		gray!30,
		gray!30,
		gray!30,
	},
	sequence item border color=black,
	sequence item font size=16,
}

\usepackage{float}

\usepackage{mathptmx}      
\usepackage{latexsym}

\usepackage{float}

\usepackage{color}
\usepackage{colortbl}

\usepackage{amssymb}
\usepackage{comment}
\usepackage[multiple]{footmisc}
\usepackage{amsmath}

\usepackage{lipsum}
\usepackage{pfnote}
\usepackage{multirow}
\usepackage{float}
\usepackage{pgfplots}
\usepackage{tikz}
\usepackage{xcolor}
\usetikzlibrary{patterns}
\usepackage{amstext}

\usepackage{color,soul}

\usepackage{colortbl}

\PassOptionsToPackage{hyphens}{url}\usepackage{hyperref}

\usepackage{amsmath}

\usepackage{enumitem}

\usepackage{colortbl}

\definecolor{maroon}{cmyk}{0,0.87,0.68,0.32}

\usetikzlibrary{positioning,calc}

\usepackage{tikz}
\usetikzlibrary{trees}

\usepackage{kbordermatrix}

\usetikzlibrary{positioning,decorations.text}

\tikzset{level 1/.style={level distance=2cm, sibling distance=10cm}}
\tikzset{level 2/.style={level distance=2cm, sibling distance=3cm}}

\tikzset{bag/.style={text width=10em, text centered,yshift=-0.5cm}}

\usepackage{tikz}
\usetikzlibrary{trees}

\tikzset{level 1/.style={level distance=2cm, sibling distance=10cm}}
\tikzset{level 2/.style={level distance=2cm, sibling distance=3cm}}

\tikzset{bag/.style={text width=10em, text centered,yshift=-0.5cm}}

\usepackage{smartdiagram}
\usesmartdiagramlibrary{additions}

\usetikzlibrary{positioning,calc}

\def\addlegendimage{\csname pgfplots@addlegendimage\endcsname}

\begin{document}

	\begin{frontmatter}
		
		
		
		\title{Exploring Diseases and Syndromes in Neurology Case Reports from 1955 to 2017 with Text Mining}
		
		\author[label1]{Amir Karami\footnote{Amir Karami will handle correspondence at all stages of refereeing and publication. \\ \textbf{Address:} Davis College, 1501 Greene St, Columbia, SC 29208. \\ \textbf{Email addresses}: karami@sc.edu (A. Karami), m82.ghasemi@gmail.com (M. Ghasemi), souvik.sen@uscmed.sc.edu (S. Sen), mmoraes@email.sc.edu (M. F. Moraes),  vshah@email.sc.edu (V. Shah)}}
		\author[label2]{Mehdi Ghasemi}
		\author[label3]{Souvik Sen}
		\author[label4]{Marcos F Moraes}
		\author[label4]{Vishal Shah}
		\address[label1]{College of Information and Communications, University of South Carolina}
		\address[label2]{Department of Neurology, Brigham and Women's Hospital, Harvard Medical School}
		\address[label3]{Neurology Department, School of Medicine, University of South Carolina}
		\address[label4]{Computer Science and Engineering Department, University of South Carolina}
		
		



\begin{abstract}
\textbf{Background}: A large number of neurology case reports have been published, but it is a challenging task for human medical experts to explore all of these publications. Text mining offers a computational approach to investigate neurology literature and capture meaningful patterns. The overarching goal of this study is to provide a new perspective on case reports of neurological disease and syndrome analysis over the last six decades using text mining. 

\textbf{Methods}: We extracted diseases and syndromes (DsSs) from more than 65,000 neurology case reports from 66 journals in PubMed over the last six decades from 1955 to 2017. Text mining was applied to reports on the detected DsSs to investigate high-frequency DsSs, categorize them, and explore the linear trends over the 63-year time frame. 

\textbf{Results}: The text mining methods explored high-frequency neurologic DsSs and their trends and the relationships between them from 1955 to 2017. We detected more than 18,000 unique DsSs and found 10 categories of neurologic DsSs. While the trend analysis showed the increasing trends in the case reports for top-10 high-frequency DsSs, the categories had mixed trends.

\textbf{Conclusion}: Our study provided new insights into the application of text mining methods to investigate DsSs in a large number of medical case reports that occur over several decades. The proposed approach can be used to provide a macro level analysis of medical literature by discovering interesting patterns and tracking them over several years to help physicians explore these case reports more efficiently.

\end{abstract}			
			
	\end{frontmatter}

	\section{Introduction}
	\label{Int}

Medical publications have been considered as primary sources to report findings in different fields such as neurology. The first line for reporting clinical evidence has been in medical case reports that contain single or multiple cases. Medical case reports (MCRs) usually represent rare cases that are the only evidence before published trials \cite{albrecht2005case}. MCRs are ``a formal summary of a unique patient and his or her illness, including the presenting signs and symptoms, diagnostic studies, treatment course and outcome" \cite{venes2009taber}, and are sometimes the first reportage of new discoveries \cite{nissen2014clinical}. However, there is no standard format for MCRs and it is difficult to get a macro level perspective \cite{nissen2014clinical}. Moreover, MCRs have been historically considered as a cost- and time-effective method in advancing knowledge and education from both patient and physician perspectives with direct impact on clinical practice and patient care \cite{sudhakaran2014role}. They have several benefits including the presentation of new applications, discussion of new diseases and syndromes (DsSs), generation of hypotheses and novel solutions, and providing educational value to the medical professions \cite{nissen2014clinical}. In the late 1990s, interest in MCRs grew. For example, the Lancet introduced a peer-reviewed ``Case Reports" section and some new journals were launched for the sole purpose of publishing MCRs \cite{nissen2014clinical}.

The increasing number of published neurology case reports makes the macro level analysis almost impossible because it is a challenge for researchers to keep up-to-date with published medical case reports. Moreover, the number of case series is far fewer than the number of single case reports \cite{mason2001case}. This issue requires a new approach to review neurology case reports with respect to DsSs entities. Our objective analysis detects high-frequency DsSs and the relationships between DsSs to overcome the generalization barrier in exploring MCRs. In addition, analyzing numerous MCRs is an opportunity for comparison and new hypothesis generation. 

Thus, in the present study, we tried to reduce the gap between the traditional case report analysis and qualitative-quantitative research methods. There are more than 2 million case reports in the PubMed database \footnote{\url{https://www.ncbi.nlm.nih.gov/pubmed?term=\%22case\%20reports\%22\%5BPublication\%20Type\%5D}}, but this database and other similar platforms don't provide data analytics features.  There are several MCR surveys that consider a small sample of publications, such as analysis of 100 cases in dermatology to evaluate the content quality of the cases \cite{albrecht2009survey}, exploration of 435 dental cases to detect bias \cite{oliveira2006critical}, review of 130 drug safety cases \cite{chakra2010case}, and review of 106 published case reports in \textit{The Lancet} \cite{albrecht2005case}. 

In neurology, traditional case series analysis has been developed to investigate specific issues such as analyzing 100 antibodies cases to investigate the effects of antibodies on N-methyl-D-aspartate (NMDA) receptors \cite{dalmau2008anti}, examining the frequency of associated tumors and concurrent neuronal autoantibodies on 20  patients \cite{hoftberger2013encephalitis}, investigating 28 case series with total of 177 patients published between 2000 and 2007 to examine the relationship of pathological gambling, dopamine agonists, and  dopamine dysregulation syndrome \cite{gallagher2007pathological}, analyzing the accuracy of the factors applied in diagnosing Alzheimer's disease for 134 patients with memory complaints \cite{lim1999clinico}, detecting the causes of cerebral palsy and their relative frequency in 217 cases between 1991 and 2001 \cite{shevell2003etiologic}, characterizing 1377 pediatric patients with intracranial aneurysms between 1991 and 2004 \cite{huang2005intracranial}, and developing a survey with 99 cases to determine the proportion of patients using tracking technologies \cite{mcshane1998feasibility}.

Analyzing a large number of case reports is a time-consuming and labor-intensive process. Therefore, applying text mining methods to identifying patterns from corpora offers promise. Among different text mining methods, topic modeling has been considered in a wide range of applications. The goal of topic modeling is to detect and group related words for the purpose of organizing and understanding documents in a corpus. This type of computational analysis is an effective and efficient approach to get a macro level view of medical case reports. Analysis of case report entities such as DsSs can provide a granular perspective of the case report studies. To fill the gap in the related literature landscape, we investigated the DsSs in the neurology case reports.

In the present study, we applied text mining and trend analysis the neurology case reports to (1) investigate high-frequency DsSs; (2) detect and categorize DsSs; and (3) explore the overall and yearly trends of the high frequency neurologic DsSs and their categories during this research time frame.  This study has research and educational applications for the experts who are interested in exploring neurological DsSs and their relationships in medical case reports.

The contributions of this paper are three-fold. First, the proposed framework offers a fully functional and automated approach to investigate a large number of the existing medical case reports and their overall trends during six decades. Research managers and policymakers can utilize our approach to understand the landscape of the neurology case reports. Secondly, the objective nature and the publicly available data of this research makes it easier to replicate the results using the shared data. Thirdly, this is the first research that finds the interactions between DsSs and discloses the patterns in a dataset containing rare medical cases.
	
\section{Methods}
	\label{ME}
	
\begin{figure}[H]
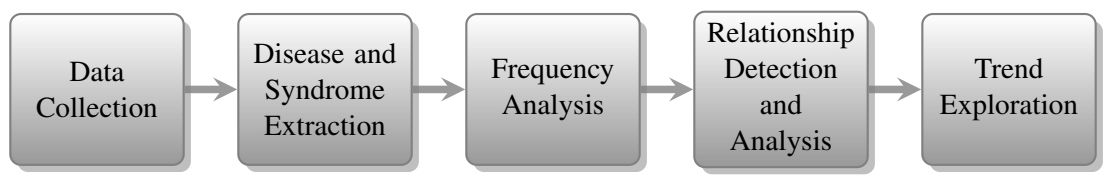

	
	\begin{center}
		\smartdiagramset{uniform color list=white!60!black for 6 items,
			back arrow disabled=true, module minimum width=2cm,
			module minimum height=2cm,
			module x sep=3cm,
			text width=2cm,
			additions={
				additional item offset=0.5cm,
				additional item width=2cm,
				additional item height=2cm,
				additional item text width=3cm
			}
		}
		\smartdiagram[flow diagram:horizontal]{Data Collection, Disease and Syndrome Extraction, Frequency Analysis, Relationship Detection and Analysis,Trend Exploration } 
	\end{center}
	\caption{Research Framework}
	\label{fig:fram} 
\end{figure}

Our approach uses text mining for disclosing the main themes of neurology case reports. The present study included five phases: data collection, disease and syndrome extraction, frequency analysis, relationship detection and analysis, and trend exploration (Figure \ref{fig:fram}). 

\subsection{Data Collection}
Data collection in this research was carried out in two steps. In the first step, we retrieved a list of top 200 neurology journals based on impact factor indicator from the Scimago Journal \& Country Rank website\footnote{\url{https://www.scimagojr.com/journalrank.php?category=2728}} in April 2018. In the second step, we retrieved the case reports based on the journal name indexed in the PubMed website. For example, we used \textit{``case reports [Publication Type] AND lancet neurol[Journal]"} query to retrieved case reports in \textit{The Lancet Neurology}. Of the 200 top neurology journals, we focused on 66 journals that published at least 50 case reports, defining them as  the highly active journals for publishing case reports  (\ref{appendix}). The collected data is available at \url{https://github.com/amir-karami/MedicalCaseReport-Diseases}.

\subsection{Disease and Syndrome Extraction}
In this step, we used PubTator\footnote{\url{https://www.ncbi.nlm.nih.gov/CBBresearch/Lu/Demo/PubTator/}} developed by the National Center for Biotechnology Information (NCBI) to extract and annotate the collected case reports \cite{wei2013pubtator}. The PubTator service can annotate diseases, species, chemicals, genes, and mutations in the PubMed documents \cite{zhu2018understanding}. Figure \ref{fig:pubtator_example} shows an example of entity detection using PubTator including title, abstract, gene, chemical, disease, and mutation. This case describes a patient having breast cancer along with gene, chemical, and mutation information. For this study, we considered the disease entity that covers both disease and syndrome (DS). 

\begin{figure}[ht]
	
	\begin{center}
		\scriptsize
		\begin{tabular}{ |p{15cm}|} 
			\hline
			29702197$\vert$t$\vert$Mutations in the estrogen receptor alpha hormone binding domain promote stem cell phenotype through notch activation in breast cancer cell lines.
			
			29702197$\vert$a$\vert$The detection of recurrent mutations affecting the hormone binding domain (HBD) of estrogen receptor alpha (ERa/ESR1) in endocrine therapy-resistant and metastatic breast cancers has prompted interest in functional characterization of these genetic alterations. Here, we explored the role of HBD-ESR1 mutations in influencing the behavior of breast cancer stem cells (BCSCs), using various BC cell lines stably expressing wild-type or mutant (Y537   N, Y537S, D538G) ERa. Compared to WT-ERa clones, mutant cells showed increased CD44+/CD24- ratio, mRNA levels of stemness genes, Mammosphere Forming Efficiency (MFE), Self-Renewal and migratory capabilities. Mutant clones exhibited high expression of NOTCH receptors/ligands/target genes and blockade of NOTCH signaling reduced MFE and migratory potential. Mutant BCSC activity was dependent on ERa phosphorylation at serine 118, since its inhibition decreased MFE and NOTCH4 activation only in mutant cells. Collectively, we demonstrate that the expression of HBD-ESR1 mutations may drive BC cells to acquire stem cell traits through ER/NOTCH4 interplay. We propose the early detection of HBD-ESR1 mutations as a challenge in precision medicine strategy, suggesting the development of tailored-approaches (i.e. NOTCH inhibitors) to prevent disease development and metastatic spread in BC mutant-positive patients.

29702197 \quad	681	\quad 685 \quad	CD24 \quad	Gene \quad	100133941

29702197 \quad	254 \quad	257 \quad	ERa	 \quad Gene	 \quad 2099

29702197 \quad	258 \quad	262	\quad ESR1	\quad Gene	\quad 2099

29702197 \quad	442 \quad	446	\quad ESR1 \quad	Gene \quad	2099

29702197 \quad	613 \quad	616	\quad ERa	\quad Gene	\quad 2099

29702197	\quad 633	\quad 636	\quad ERa	\quad Gene	\quad 2099

29702197	\quad 991	\quad 994	\quad ERa	\quad Gene	\quad 2099

29702197	\quad 1161	\quad 1165 \quad	ESR1 \quad	Gene\quad	2099

29702197\quad	1290\quad	1294\quad	ESR1\quad	Gene\quad	2099

29702197\quad	1234\quad	1240\quad	NOTCH4\quad	Gene\quad	4855

29702197\quad	1014\quad	1020\quad	serine\quad	Chemical\quad	CHEBI:17822

\textbf{29702197\quad	310	\quad 324 \quad	breast \quad cancers\quad	Disease	\quad D001943}

\textbf{29702197\quad	488\quad	501\quad	breast cancer\quad	Disease	\quad D001943}

29702197	\quad 1501\quad	1509\quad	patients\quad	Species\quad	9606

29702197\quad	599	\quad 604\quad	Y537S\quad	Mutation\quad	p $\vert$  SUB$\vert$Y$\vert$537$\vert$S \\ \hline
		\end{tabular}
	\caption{PubTator Example}
	\label{fig:pubtator_example}
	\end{center}
\end{figure}

\subsection{Frequency Analysis}			
Frequency analysis provides overall information about the detected DsSs \cite{van2016understanding}. Exploring high-frequency DsSs with bar chart and word cloud is a starting point for content analysis. Word cloud analysis visually summarizes text analysis and provides an overall view of the data in a corpus. This method presents the frequency of words in a corpus where the word size is proportional to the frequency of the word within the corpus- the larger size, the higher frequency. This approach has been used for a variety of applications, such as opinion mining  \cite{heimerl2014word}. Using word clouds to explore high-frequency DsSc reveals the important DsSc in the case reports over the past decades.

\subsection{Relationship Detection and Analysis}
\begin{figure}[htp!]
	\centering
	\includegraphics[width=0.8\textwidth]{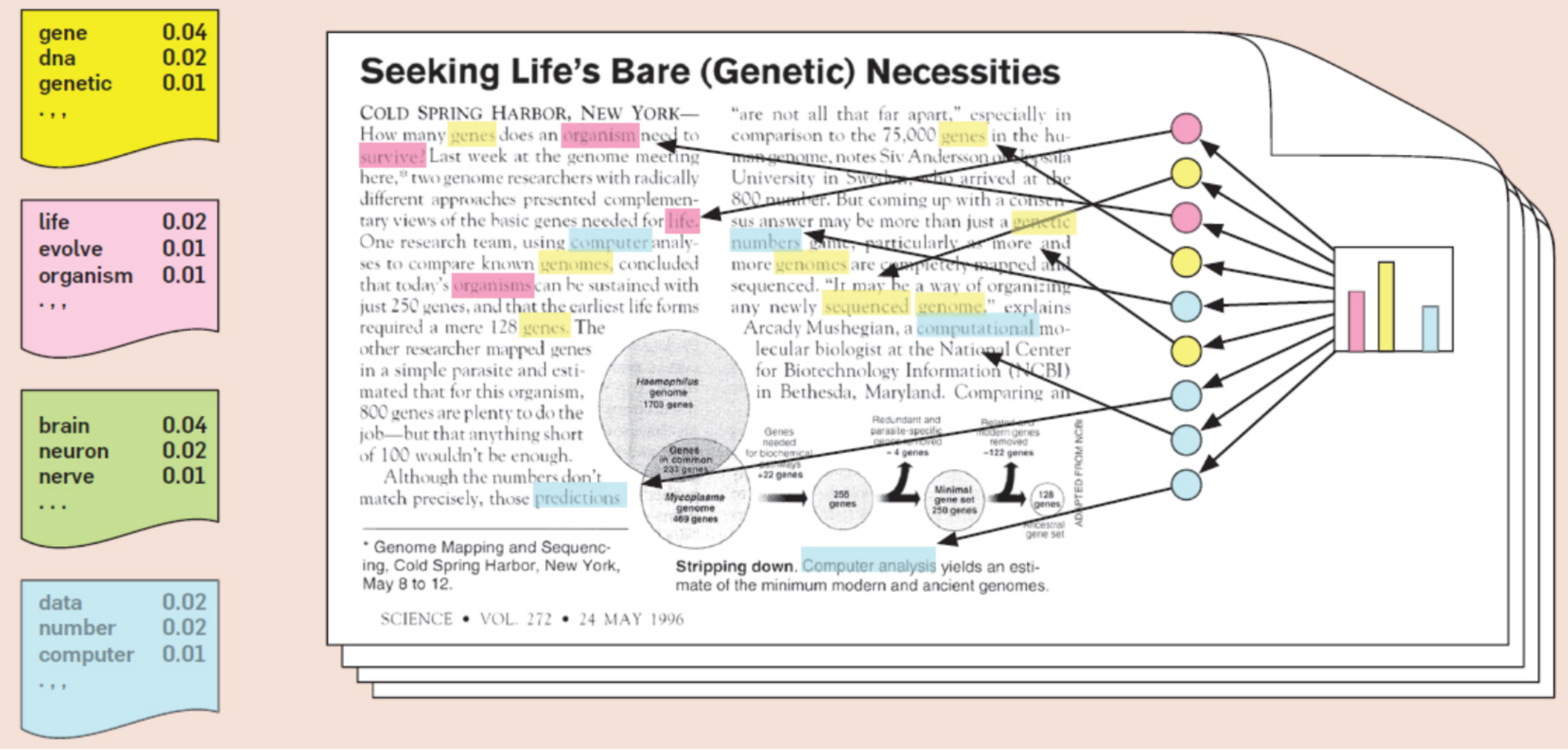}
	\caption{An Example of LDA \cite{blei2012probabilistic}}
	\label{fig:LDA_Example}
\end{figure}

Next we turned our attention to detect relationships between DsSs. The goal of this analysis was to find DsSs that were discussed together in multiple case reports. There are two major approaches in content analysis for detecting the relationships between words in a corpus: co-occurrence analysis and the afore-mentioned  topic modeling \cite{sugimoto2011shifting}. Previous studies have shown that topic modeling shows a better performance than co-occurrence analysis for large corpora \cite{leydesdorff2017co}. Topic modeling has been applied on both short-length documents such as tweets and long-length documents like research papers \cite{karami2015fuzzyiconf,karami2015flatm}. Among different topic models, latent Dirichlet allocation (LDA) is the most popular topic model and also has demonstrated better performance than co-occurrence analysis with respect to word clustering \cite{karami2015fuzzy,lu2012measuring}. According to the literature, LDA is a valid and widely used model with more than 25,000 citations in Google Scholar\footnote{\url{https://scholar.google.com/scholar?cites=17756175773309118945\&as\_sdt=5,41\&sciodt=0,41\&hl=en}} for discovering categories of words in a corpus. LDA is a generative probabilistic model for categorizing the words that occur together in a corpus \cite{blei2003latent,karami2018fuzzy}. For example in Figure \ref{fig:LDA_Example}, LDA assigns ``gene,"  ``dna," and ``genetic" into the same category. 

LDA has been utilized for health applications such as diet, diabetes, exercise, and obesity \cite{karami2019exploratory,karami2018characterizing,shaw2017computational}, and LGBT health issues \cite{karami2018characterizingtrans,webb2018Characterizing}, and non-health applications such as business and organizations \cite{collins2018social,karami2018us,karami2018computational}, spam detection \cite{karami2014exploiting}, disaster management \cite{karami2019twitter}, and politics \cite{karami2019political,karami2018mining}. There are some work investigating related studies in medical and health domains using LDA such as exploring the literature of depressive disorders \cite{zhu2018understanding}, biomedical  literature \cite{chen2017revealing,van2016understanding}, and adolescent substance use and depression \cite{wang2016text}.

To the best of our knowledge, this study is the first research uses LDA to analyze medical case reports. LDA assumes that there is an exchange between words and documents in a corpus \cite{karami2018fuzzy}. In this research, we represent each case report (document) with the embedded DsSs (words) and don't consider other words such as title and abstract in the case report. Based on the LDA assumption, the neurology case reports are represented by categories of DsSs in our dataset. Another assumption of LDA is that a category, which is a group of related DsSs, has a different probability of occurrence for each of the case reports. This assumption can help to measure the weight of categories in each case report. In summary, LDA identifies the relationship between case reports and the categories, $P(Category|CaseReport)$, and categories and DsSs, $P(DS|Category)$.

For $n$ case reports, $m$ DsSs, and $t$ categories, the outputs of LDA were: the probability of each of the DsSs per each category or $P(DS_i|C_k)$ and probability of each of the categories per each case report or $P(C_k|CR_j)$:

\begin{figure}[H]
	\small
	\[
	LDA \rightarrow 
	\kbordermatrix{
		& & Categories& \\
		&P(DS_1|C_1) & \dots & P(DS_1|C_t)\\
		DsSs &\vdots & \ddots & \vdots\\
		&P(DS_m|C_1) &   \dots     & P(DS_m|C_t) 	
	}
	\&
	\kbordermatrix{
		& & Case Reports& \\
		&P(C_1|CR_1) & \dots & P(C_1|CR_n)\\
		Categories&\vdots & \ddots & \vdots\\
		&P(C_t|CR_1) &   \dots     & P(C_t|CR_n) 	
	}
	\]
\end{figure}

The top DsSs in each category, based on the order of $P(DS_i|C_k)$, were used to represent the categories. We also used $P(C_k|CR_j)$ to find the \textit{\underline{S}}ignificance of each \textit{\underline{C}}ategory, $SC(C_k)$. For an effective comparison between the categories, $SC(C_k)$ was normalized: 

\begin{center}
$N\_SC(C_k) = \frac{\sum_{j=1}^{n} P(C_k|CR_j)}{\sum_{k=1}^{t} \sum_{j=1}^{n} P(C_k|CR_j)}$
\end{center}

If $N\_SC(C_x) > N\_SC(C_y)$, it means that researchers discussed the diseases in category x more than the ones in category y. $N\_SC(C_k)$ can also help to find the weight of each category for each year and all the years.

\subsection{Trend Exploration}
We explored the DsSs trends using a linear trend model across six decades. We used the frequency of DsSs and the categories of DsSs in the case reports within each year to detect and identify increasing and decreasing trends in research interest around the DsSs \cite{zhu2018understanding}. We used the \textit{lm} function in R to measure $p-value$ and slope for each of the DsSs and categories. The $p-value$ determines whether a trend is significant and the slope shows whether a trend is increasing and decreasing. The trend shows increasing or decreasing importance of an entity. The slope shows the intensities of the trends, useful in comparing them.

\section{Results}
\label{Ex}

\begin{figure}[htp!]
	\centering
	\includegraphics[width=0.5\textwidth]{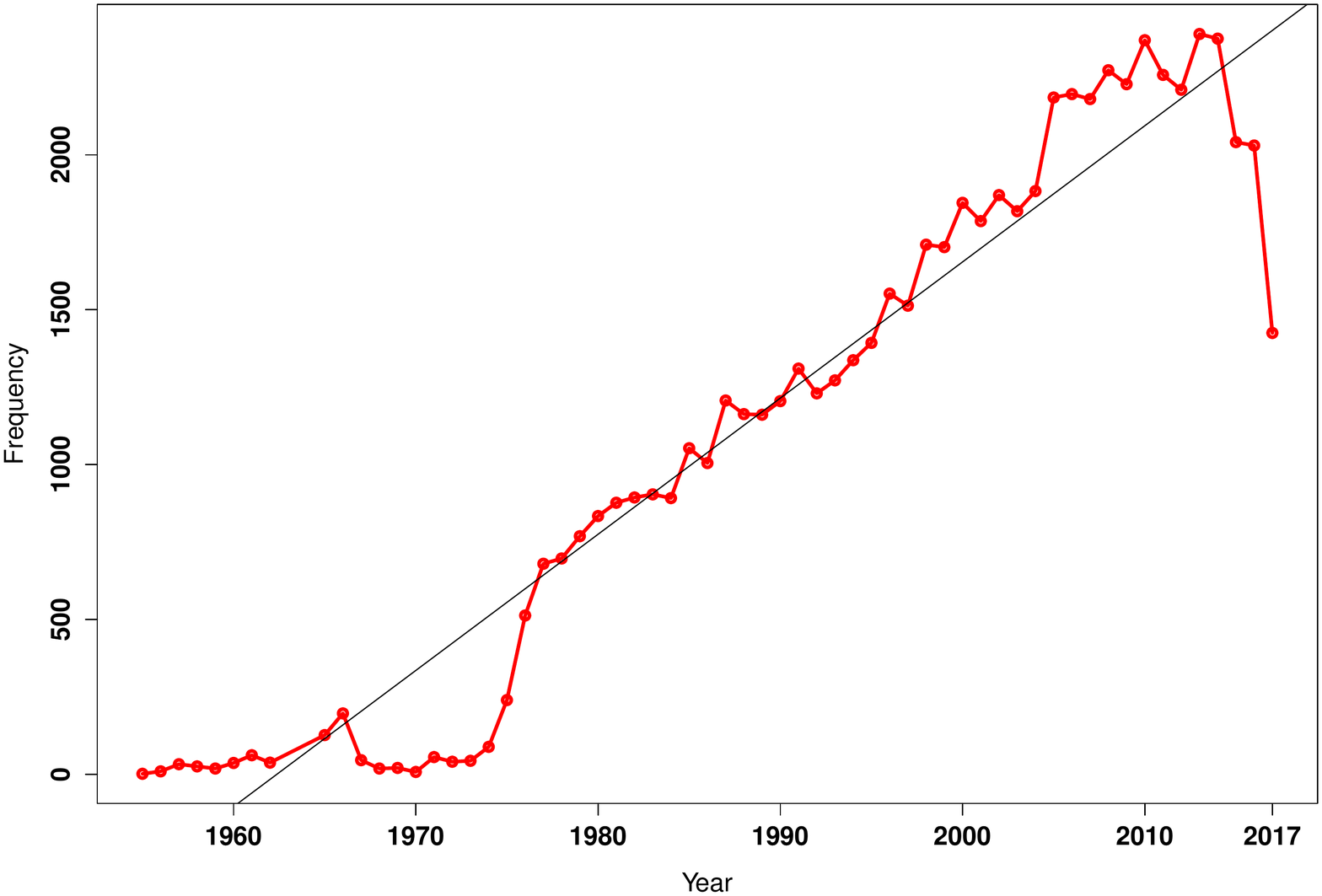}
	\caption{Frequency of case reports from 1955 to 2017}
	\label{fig:CaseReports_Year_till_2017}
\end{figure}

The data collection step has provided 65,525 case reports for 63 years, from 1955 to 2017, in the MEDLINE format.  Figure \ref{fig:CaseReports_Year_till_2017} shows the number of published neurology case reports per year over more than six decades along with the trend line. The linear trend was significant ($p<0.05$) and had a positive value (43.95) for the slope indicating an increasing trend. 

\begin{figure}[H]
	\centering
	\includegraphics[width=0.5\textwidth]{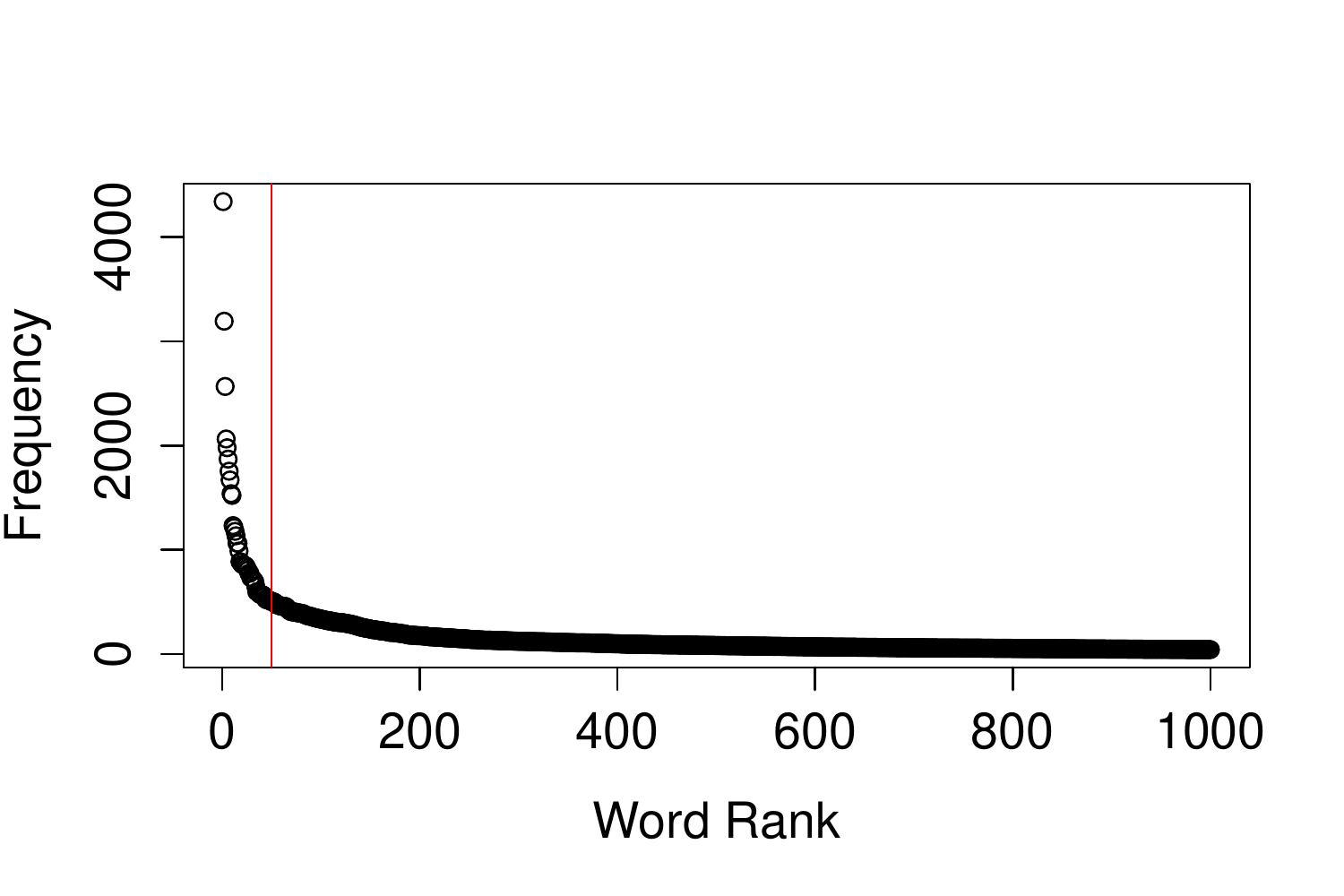}
	\caption{Frequency of DsSs. The vertical line shows the cut-off point for top-50 words in the word cloud.}
	\label{fig:WordRank_Frequency_Figure_1000}
\end{figure}

PubTator detected DsSs in 93\% (61,149 out of 65,625) of the case reports. We found 258,487 DsSs among which 18,081 DsSs were unique. Word frequency analysis showed that 95\% of DsSs occurred fewer than 50 times. With a median of 3 and an average of 14.3, the frequency of DsSs was between 2 (for 6,675 of the DsSs) and 4,319 (for tumor). Figure \ref{fig:WordRank_Frequency_Figure_1000} is in line with Zipf's law and illustrates the position of the top 50 words among the top 1000 high-frequency words. Zipf's law states that the frequency of a word in a corpus is inversely proportional to its frequency rank \cite{zipf1949human}.

\begin{figure}[htp!]
	\centering
	\includegraphics[width=0.8\textwidth]{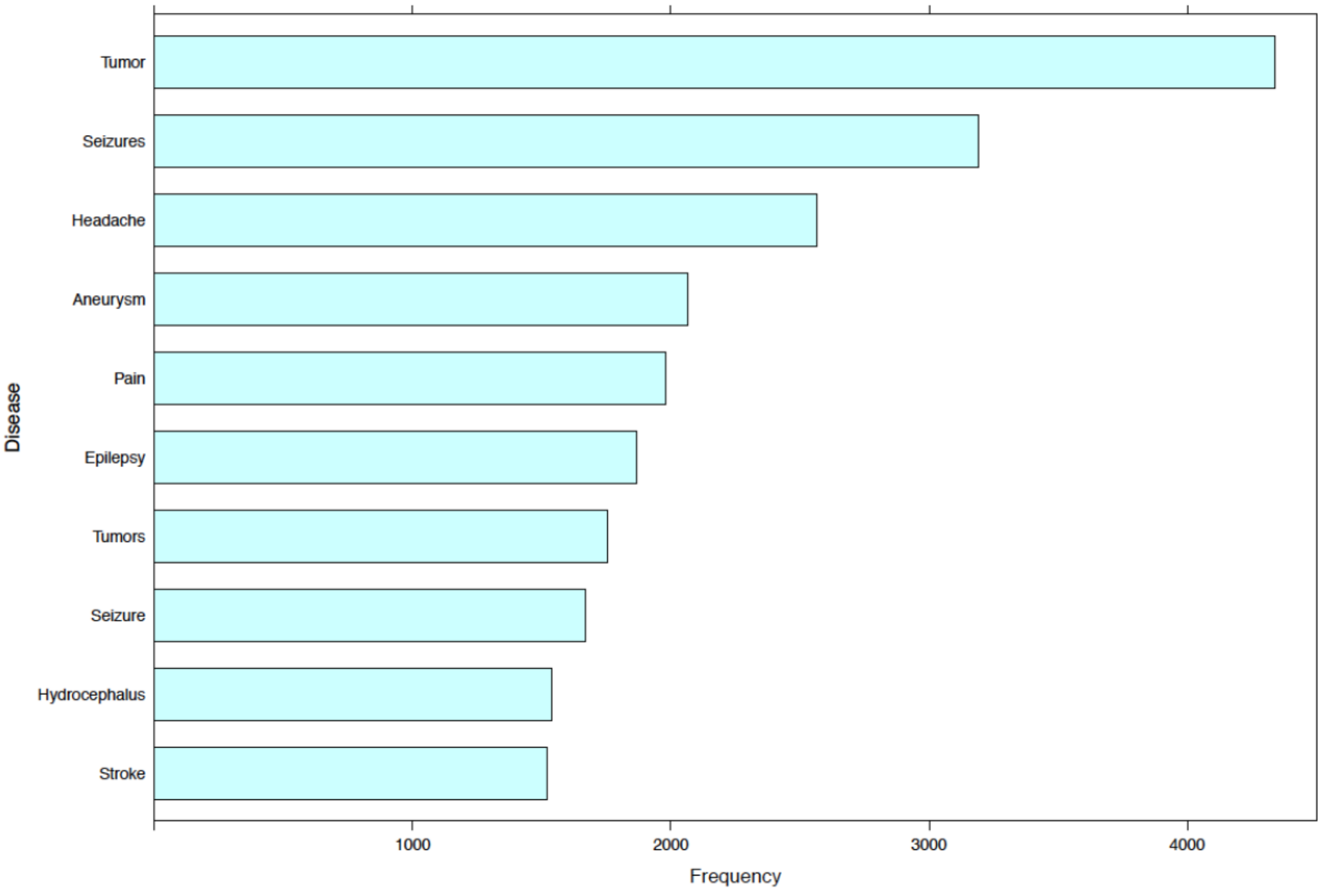}
	\caption{Frequency of the top 10 DsSs}
	\label{fig:Top_10_Diseases_BarChart_3}

	\includegraphics[width=0.8\textwidth]{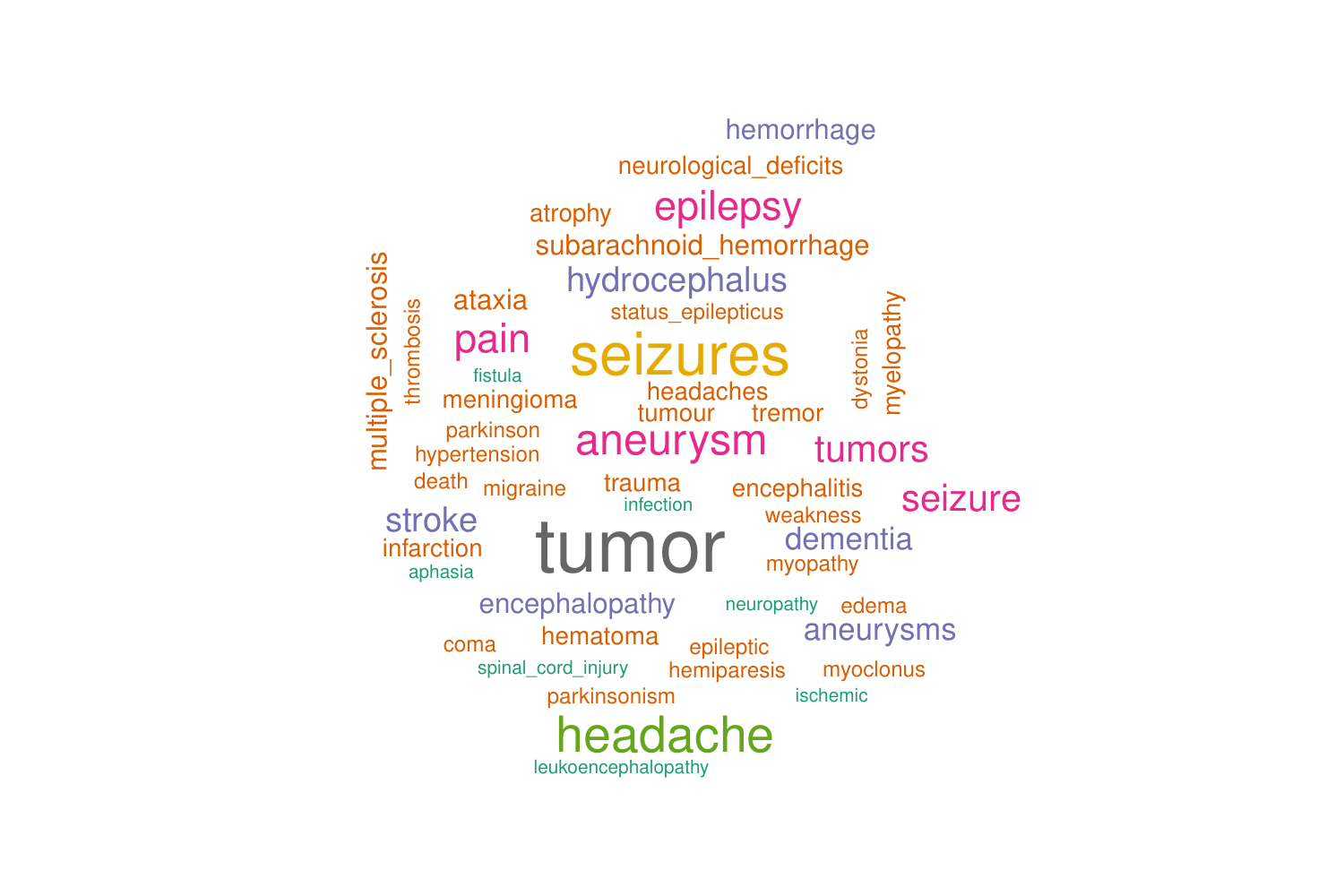}
	\caption{Word cloud of the top 50 DsSs}
	\label{fig:WordCloud_Top_50_Words}
\end{figure}

Figures \ref{fig:Top_10_Diseases_BarChart_3} and \ref{fig:WordCloud_Top_50_Words} illustrate the frequency of the top 10 high-frequency DsSs and the word cloud for the top 50 DsSs. These two figures indicate that that tumor, seizures, and headache are the most frequent DsSs, characterizing 4339, 3190, and 2566 case reports, respectively. We didn't apply any stemming techniques in the identification of categories in case reports.  For example, we didn't convert a plural form such as tumors to the singular one to distinguish the case reports studying multiple tumors from the ones investigating a single tumor.

\begin{figure}[htp!]
	\centering
	\includegraphics[width=0.7\textwidth]{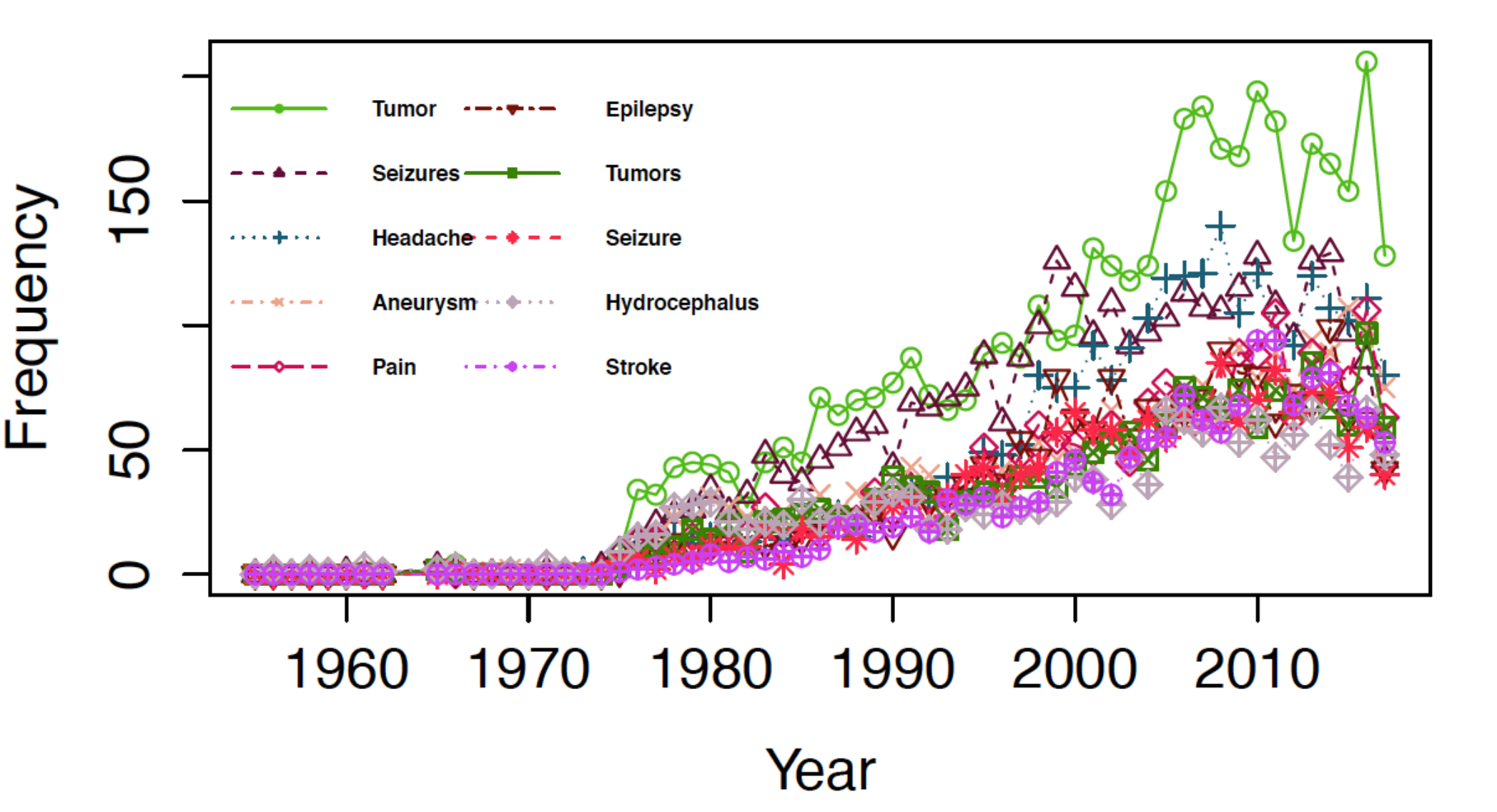}
	\caption{Overall linear trends for the top10 DsSs}
	\label{fig:Top_10_Disease_Trend_2}
\end{figure}

Figure \ref{fig:Top_10_Disease_Trend_2} shows overall trends for the top 10 DsSs. Tumor was the most frequently mentioned DsSs in the 1970s, 1980s, 2000s, and after 2010. However, seizures were the most frequent one in the 1990s. The frequency of headache was close to and in some years higher than the frequency of seizures after 2000.

\begin{figure}[htp!]
	\centering
	\includegraphics[width=0.8\textwidth]{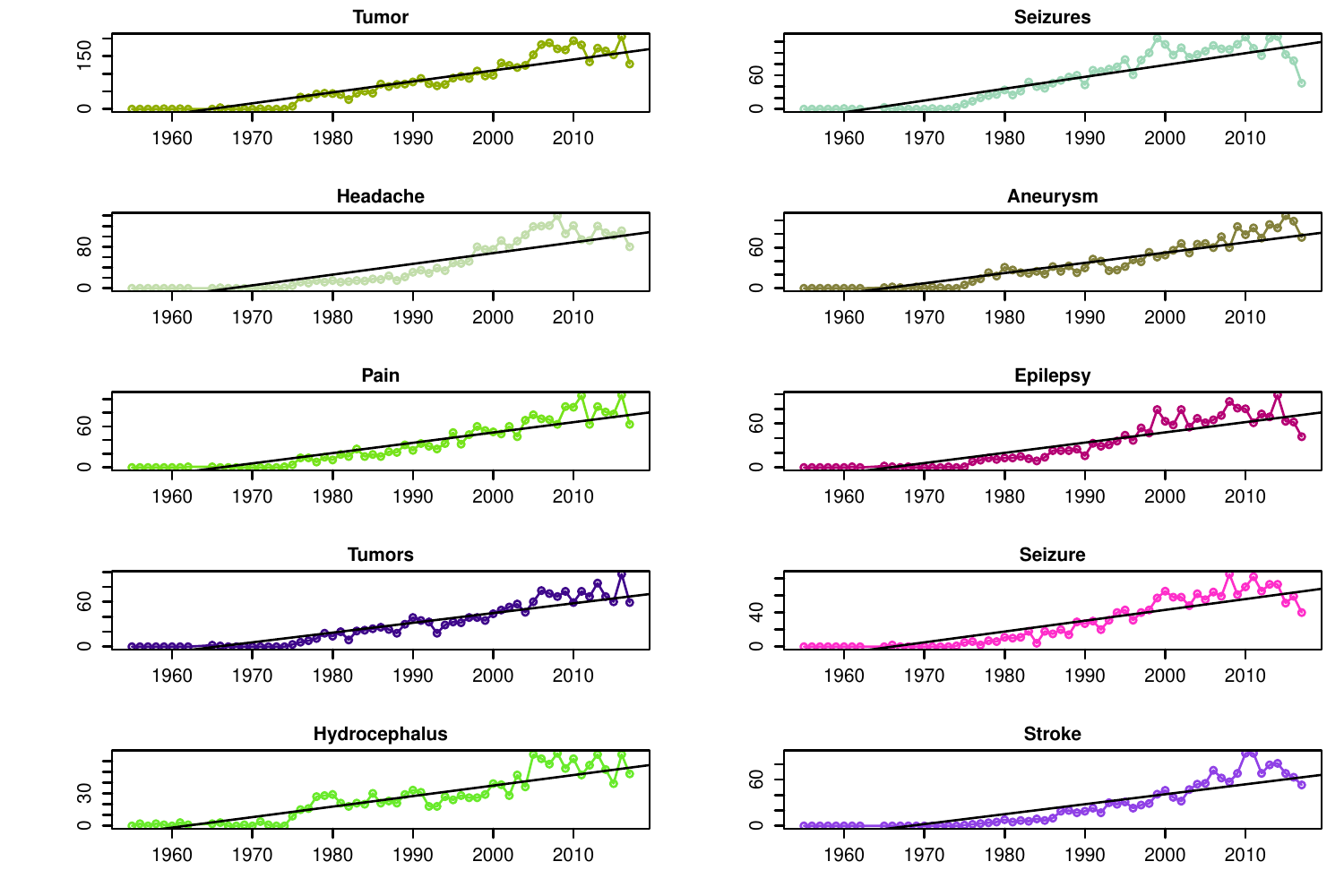}
	\caption{Linear trends of the top 10 DsSs from 1955 to 2017}
	\label{fig:Top_10_Disease_Trend_Multi_Figures}
\end{figure}

Figure \ref{fig:Top_10_Disease_Trend_Multi_Figures} and Table \ref{tab:diseases} illustrate the linear trend information of the top 10 DsSs from 1955 to 2017. The average number of studies per year for the top 10 DsSs was more than 25. The linear trend of the top 10 DsSs was significant ($p<0.05$) with a positive slope indicating increasing trends. Based on the slope values, we expected to see the same ranking and patterns for 9 out of 10 DsSs in the following years. The only exception was that the slope value of hydrocephalus was less than the slope value of stroke. This means that we could see more stroke-related case reports than hydrocephalus-related ones.

\begin{table}[htp!]
	
	\caption{Linear trend data of top-10 DsSs from 1955 to 2017}
	\begin{center}
		\begin{tabular}{  ccccc} 
			\hline
			{\cellcolor[gray]{.9}}  \textbf{Disease} & {\cellcolor[gray]{.9}}  \textbf{Frequency} & {\cellcolor[gray]{.9}}  \textbf{Average/Year} & {\cellcolor[gray]{.9}}  \textbf{Slope} & {\cellcolor[gray]{.9}}  \textbf{P-Value} \\  		\hline
			\textbf{Tumor} &	4339&	68.87&	3.11&	0.00\\  			
			\textbf{Seizures}&	3190&	50.63&	2.10&	0.00\\  			
			\textbf{Headache}&	2566&	40.73&	2.07&	0.00\\  			
			\textbf{Aneurysm}&	2066&	32.79&	1.50&	0.00\\  			
			\textbf{Pain}&	1979&	31.41&	1.50&	0.00\\  			
			\textbf{Epilepsy}&	1867&	29.63&	1.40&	0.00\\  			
			\textbf{Tumors}&	1754&	27.84&	1.30&	0.00\\  			
			\textbf{Seizure}&	1670&	26.51&	1.27&	0.00\\  			
			\textbf{Hydrocephalus}&	1539&	24.43&	0.97&	0.00\\  			
			\textbf{Stroke}&	1521&	24.14&	1.29&	0.00 \\  	\hline			
		\end{tabular}
		
		\label{tab:diseases}
	\end{center} 
\end{table}

We found that 81\% (50,944 out of 61,149) of case reports discussed multiple DsSs. This research used the Mallet implementation of LDA using Gibbs sampling with its default settings to categorize DsSs in the case reports with multiple DsSs. 

To select an optimal number of categories, we applied a density-based method that assumes that the best performance of LDA is at the minimum average cosine distance of categories \cite{cao2009density}. Applying the \textit{ldatuning} R package\footnote{\url{https://cran.r-project.org/web/packages/ldatuning/vignettes/topics.html}}  on the number of categories from 5 to 100 increased by 5 has shown the appropriate number of categories at 10. The Mallet implementation, which is a Java programming language for text mining purposes \cite{mallet}, was applied on the DsSs with 10 categories and 1000 iterations.

\begin{figure}[htp!]
	\centering
	\includegraphics[width=0.6\textwidth]{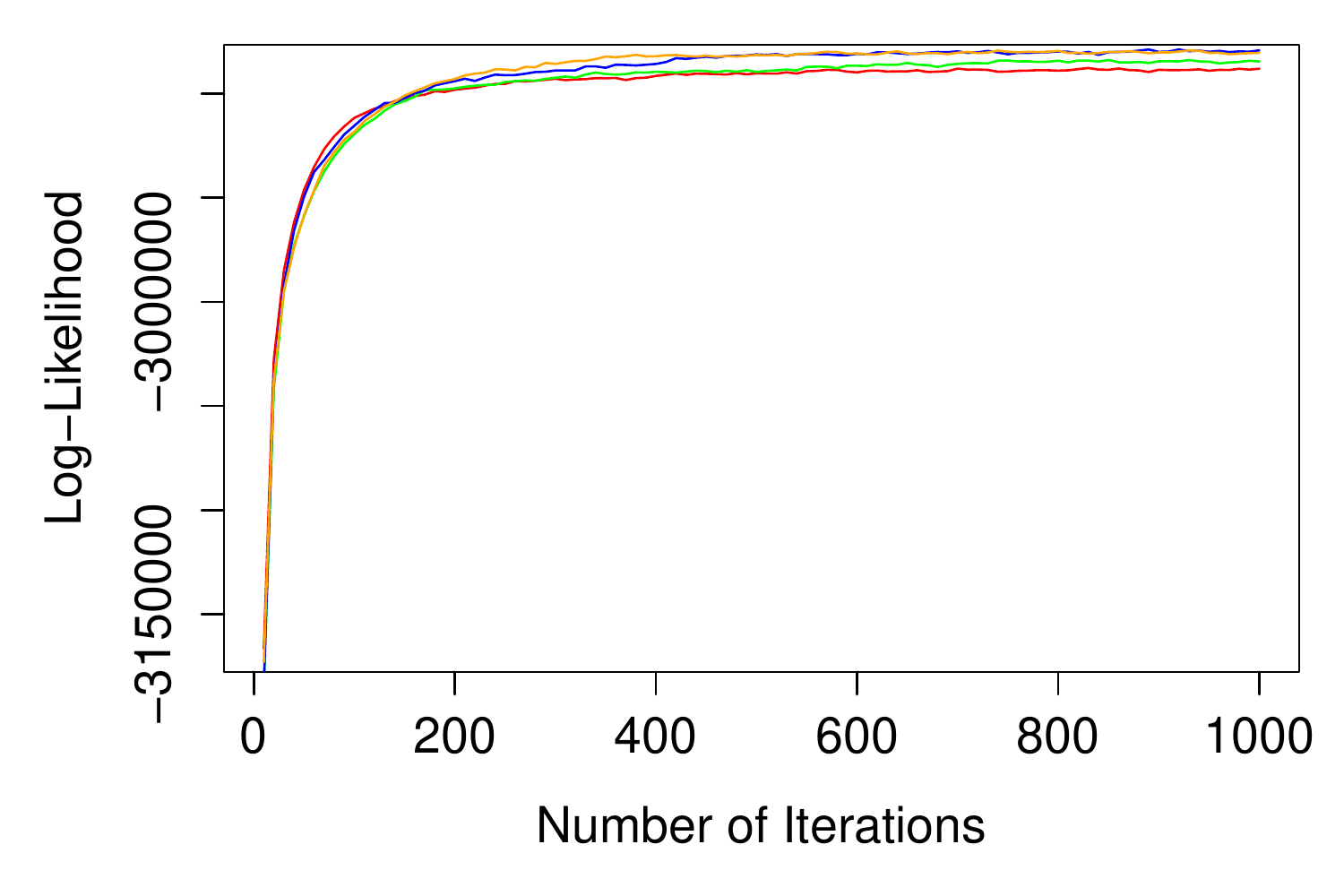}
	\caption{Convergence of the log-likelihood for 5 sets of 1000 iterations}
	\label{fig:LogLikeligood_Iterations}
\end{figure}

Then, we evaluated the robustness of LDA on using the log-likelihood for five sets of 1000 iterations. LDA was trained on the case reports having multiple DsSs with 10 categories for five times and reached its maximum value after 500 iterations (Figure \ref{fig:LogLikeligood_Iterations}). We compared the five iterations using t-test and found that there isn't a significant difference ($p-value >0.05$) between the five iterations with respect to mean and standard deviation.  This figure and the t-test show that there isn't a significant difference in log-likelihood convergence over different iterations.

\begin{figure}[htp!]
	\centering
	\includegraphics[width=0.5\textwidth]{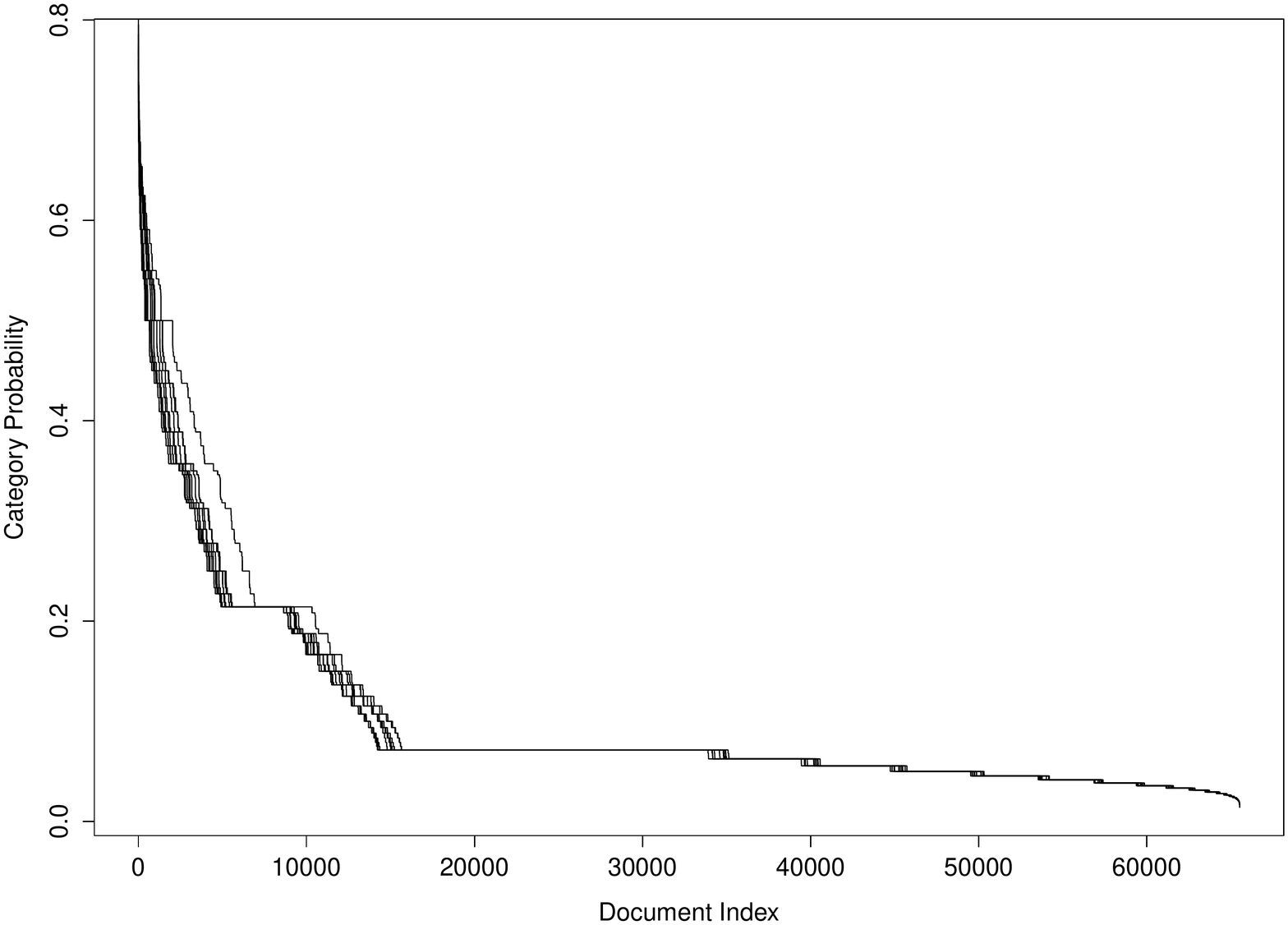}
	\caption{Distribution of categories over the case reports with multiple DsSs}
	\label{fig:TopicDocuemntDistribution_Red_Line_Topic6}
\end{figure}

Figure \ref{fig:TopicDocuemntDistribution_Red_Line_Topic6} shows the distribution of categories over the case reports and each line represents a category. This figure indicates that the 10 categories have a high probability for less than 2000 case reports.

\begin{figure}[htp!]
	\centering
	\includegraphics[width=0.6\textwidth]{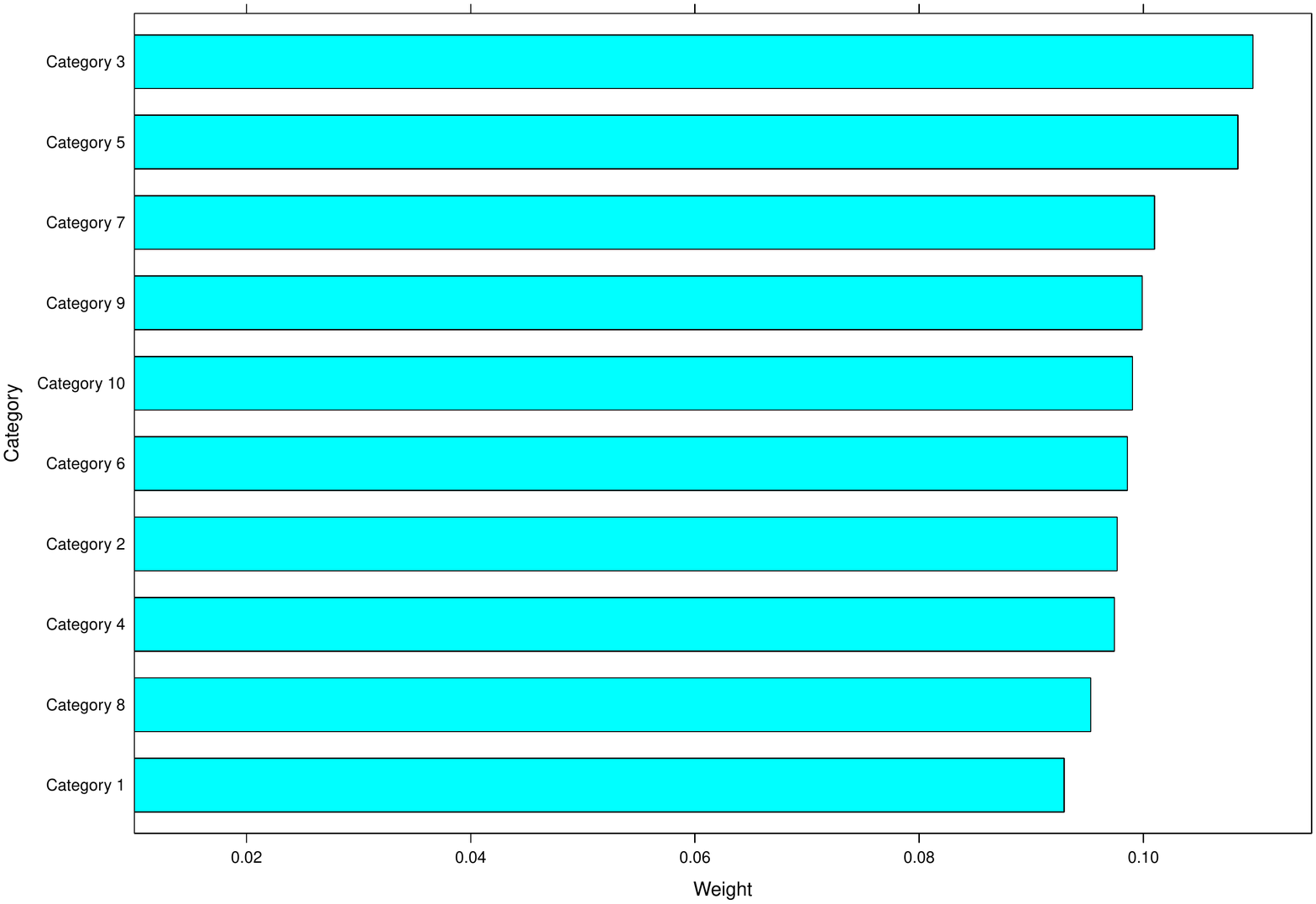}
	\caption{Weight of Categories}
	\label{fig:Popularity_Categories}
\end{figure}

Table \ref{tab:topics} shows the 10 categories. These categories show the possible relationships between the top 5 DsSs in each of the categories. We also explored the weight of categories and found that categories 3, 5, and 7 are the most discussed ones; however, the zero value of the standard deviation of the weights ranging from 0.09 to 0.11  shows no significant difference among the weights (Figure \ref{fig:Popularity_Categories}). Therefore, we can assume that the categories have similar weight.

\begin{table}[htp!]
	\centering			
	
	\caption{Categories of relevant DsSs}
	\scalebox{0.8}{
		\begin{tabular}{ ccccc}\hline
			\rowcolor[gray]{.9} \textbf{Category 1}& \textbf{Category 2} & \textbf{Category 3}  & \textbf{Category 4} & \textbf{Category 5}    \\\hline
			
			headache      & stroke           & dementia       & multiple\_sclerosis & tremor       \\
			hydrocephalus & infarction       & atrophy        & encephalitis        & dystonia     \\
			headaches     & ischemic         & ataxia         & encephalopathy      & parkinson    \\
			pain      & hemiparesis      & neuronal\_loss & meningitis          & parkinsonism \\
			migraine      & ischemic\_stroke & gliosis        & infection           & depression  \\\hline
			
			\rowcolor[gray]{.9} \textbf{Category 6}& \textbf{Category 7} & \textbf{Category 8}  & \textbf{Category 9} & \textbf{Category 10}    \\\hline
			tumor       & weakness               & pain                & aneurysm                & seizures            \\
			tumors       & neuropathy             & myelopathy           & aneurysms                & epilepsy           \\
			meningioma   & myopathy               & trauma               & hemorrhage               & seizure          \\
			tumour       & polyneuropathy         & spinal\_cord\_injury & subarachnoid\_hemorrhage & epileptic           \\
			glioblastoma & peripheral\_neuropathy & paraplegia           & hematoma                 & status\_epilepticus	\\	 \hline
			
		\end{tabular}}			
		\label{tab:topics}		
	\end{table}

\begin{figure}[htp!]
	\centering
	\includegraphics[width=0.8\textwidth]{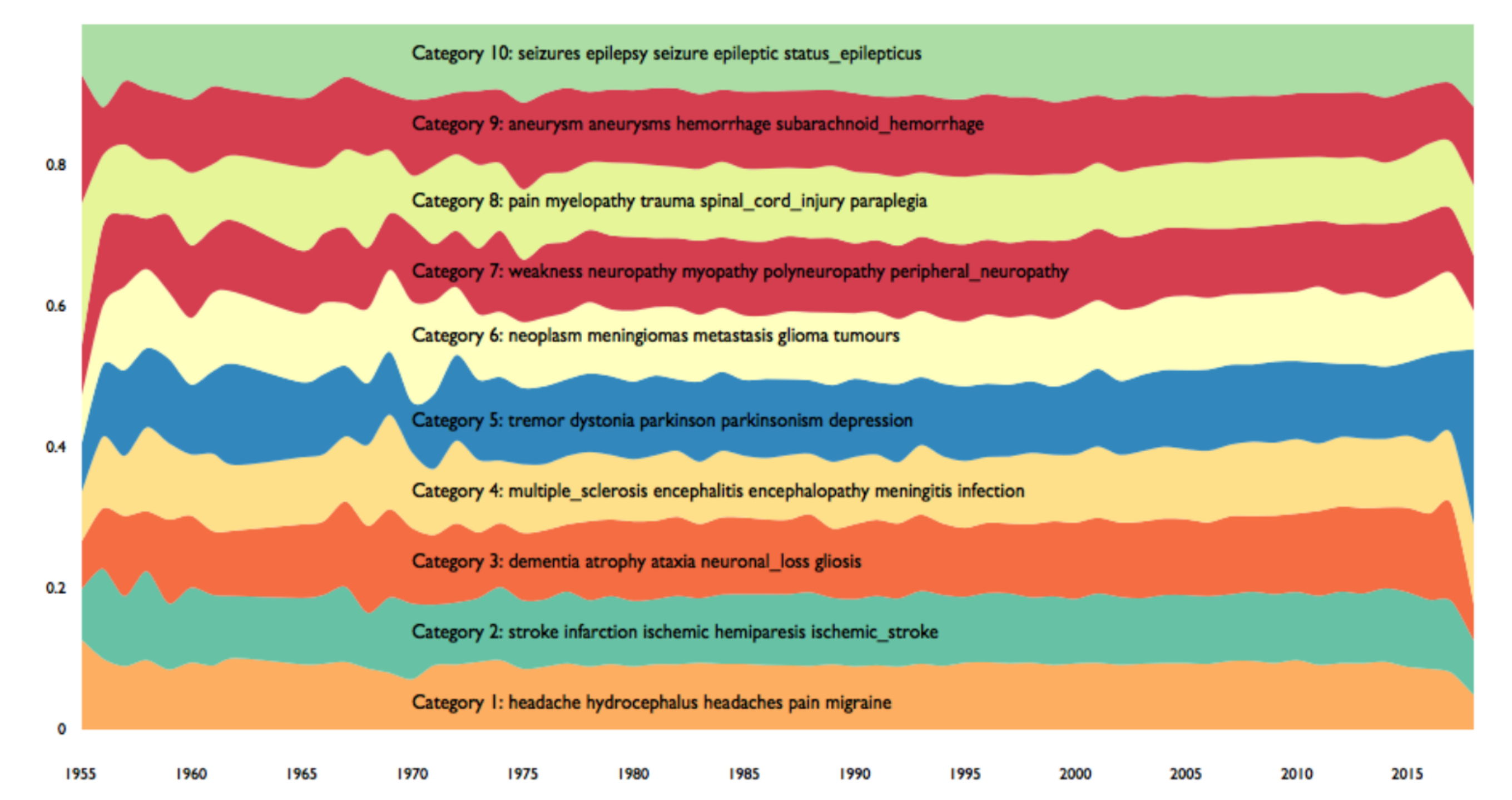}
	\caption{Yearly stream graph of categories}
	\label{fig:Categories_Trend_Stream_Years}
\end{figure}

\begin{figure}[htp!]
	\centering
	\includegraphics[width=0.8\textwidth]{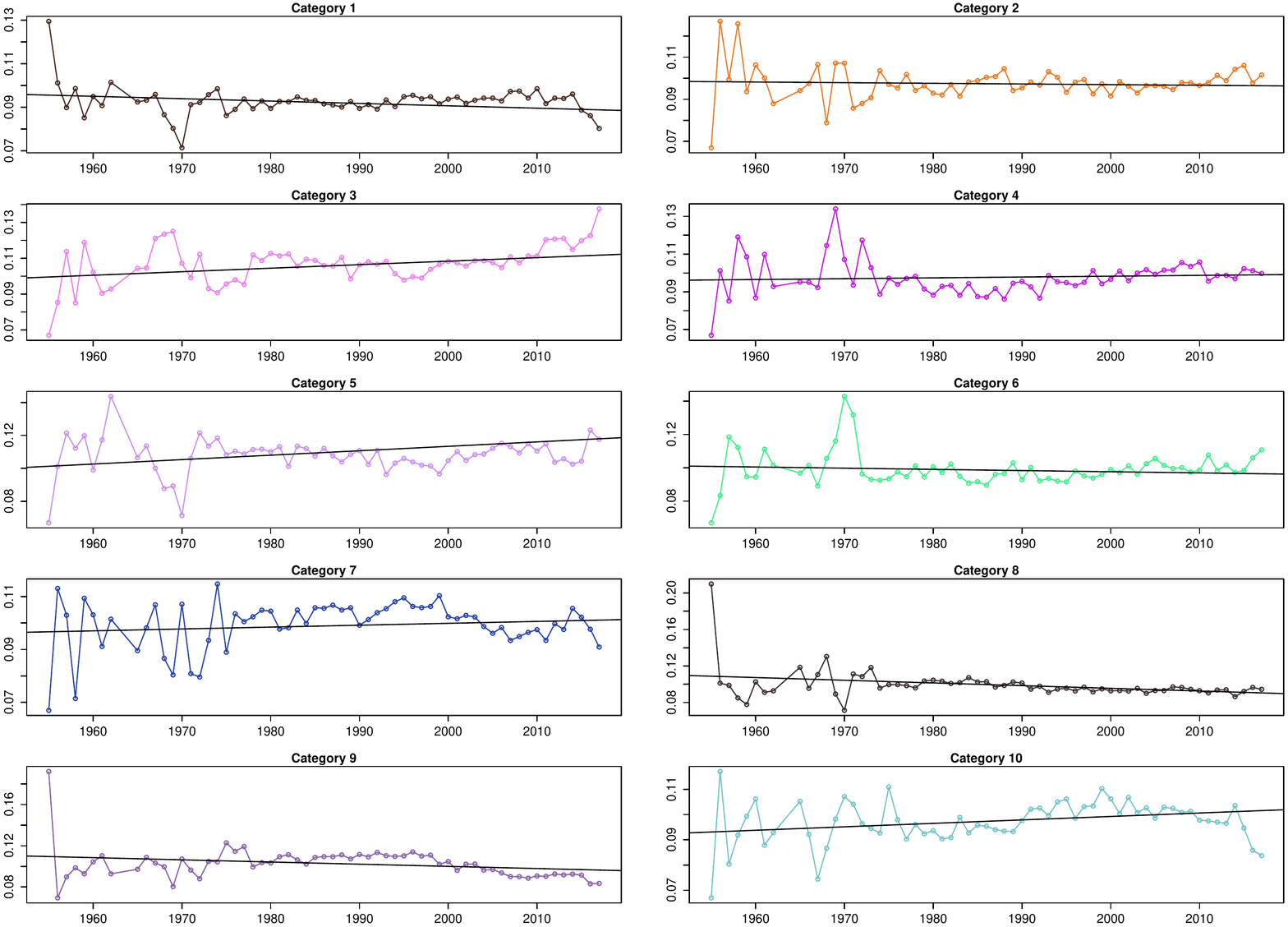}
	\caption{Yearly linear trends of DsSs}
	\label{fig:Categories_Trend_Multi_Figures}
\end{figure}

We investigated the yearly trends of categories for six decades from 1955 to 2017 (Figure \ref{fig:Categories_Trend_Stream_Years}). The stream figure visualizes the evolution of categories based on the probability of the 10 categories in each of the years.  Each color represents a category that horizontally flows from left to right. In Table \ref{tab:diseases_trend} and Figure \ref{fig:Categories_Trend_Multi_Figures}, although $p-value$ $<0.05$ shows that the trends for categories 3, 8, and 10 are significant.

	\begin{table}[htp!]
		
		\caption{Yearly Linear Trends of Diseases}
		\begin{center}
			\begin{tabular}{  cccc} 
				\hline
				{\cellcolor[gray]{.9}}  \textbf{Category} & {\cellcolor[gray]{.9}}  \textbf{Weight} & {\cellcolor[gray]{.9}}  \textbf{Slope} & {\cellcolor[gray]{.9}}  \textbf{P-Value}  \\  		\hline
				\textbf{Category 1}  & 0.0929  & -0.000109 & 0.071897 \\ 
				\textbf{Category 2}  & 0.0977  & -0.000031 & 0.612404 \\ 
				\textbf{Category 3}  & 0.1098  & 0.000197  & 0.028806 \\ 
				\textbf{Category 4}  & 0.0974  & 0.000044  & 0.512252 \\ 
				\textbf{Category 5}  & 0.1084  & 0.000269  & 0.064952 \\ 
				\textbf{Category 6}  & 0. 0985 & -0.000071 & 0.389706 \\ 
				\textbf{Category 7}  & 0.1009  & 0.000071  & 0.287937 \\ 
				\textbf{Category 8}  & 0.0953  & -0.000295 & 0.010105 \\ 
				\textbf{Category 9}  & 0.0999  & -0.000211 & 0.047734 \\ 
				\textbf{Category 10} & 0.099   & 0.000137  & 0.023201 \\ \hline	
			\end{tabular}
			
			\label{tab:diseases_trend}
		\end{center} 
	\end{table}

\newpage
	
\section{Discussion}
\label{Di}
Due to the lack of a macro level analysis and the high volume of medical case reports, there is a need to utilize computational methods for the purpose of analyzing a large number of case reports.  In this study we investigated DsSs in the neurology case reports to recognize patterns and provide additional insights using text mining. Analysis of case report entities such as DsSs provides a ``bird's eye view" to understand the importance of the entities and their interactions.

Exploring the patterns of DsSs will provide vital insights into many aspects as follows. First, the high-frequency DsSs suggest their high importance to medical experts. Second, the DsSs in a category that have not been studied together can be considered as useful hypotheses for further investigation. Third, the detected trends show research streams. Fourth, sharp changes in the detected trends provide an overview of past studies and insights for future studies.  

Through frequency analysis, we found high-frequency DsSs such as tumor(s), seizure(s), headache, aneurysm, pain, epilepsy, hydrocephalus, and stroke. This analysis shows the importance of these DsSs for researchers. Relationship detection and analysis generated 10 categories of DsSs. For example, the first category shows the case reports mentioning two or more of the represented DsSs, such as a paper reporting headache, hydrocephalus, pain, and migraine  \cite{schlake1991symptomatic}. Examples of other case reports for each of the categories can be found in Table \ref{tab:Case_Report_Examples}.

	\begin{table}[htp!]
		
		\caption{Case report examples for categories}
		\begin{center}
			\begin{tabular}{ ccc} 
				\hline
				{\cellcolor[gray]{.9}}  \textbf{Category} & {\cellcolor[gray]{.9}}  \textbf{Related Case Report} & {\cellcolor[gray]{.9}}  \textbf{Reported Diseases and Syndromes}   \\  		\hline
				1        & \cite{schlake1991symptomatic}        & headache, hydrocephalus, pain, migraine \\ 
				2        & \cite{sanefuji2006moyamoya}       & stroke, hemiparesis                     \\ 
				3        & \cite{iwasaki2013autopsied}        & ataxia, atrophy, neuron loss            \\ 
				4        & \cite{akuzawa2014secondary}        & meningitis , infection                  \\ 
				5        & \cite{haq2010case}            & tremor, dystonia, parkinson             \\ 
				6        & \cite{mathis2013intracranial}          & tumor, meningioma                       \\ 
				7        & \cite{schuitevoerder1997proximal} & weakness, myopathy                      \\ 
				8        & \cite{macdonald1988microcystic}      & myelopathy, spinal cord injury          \\ 
				9        & \cite{tumialan2008intracranial}      & aneurysm, hemorrhage                    \\ 
				10       & \cite{marras2011deep}      & epilepsy, seizure                    \\  \hline
			\end{tabular}
			\label{tab:Case_Report_Examples}
		\end{center}
	\end{table}

The stream river figure has visualized how the ten categories have changed over more than 60 years. The width variation reflects the strength of the categories at each time slice. For example, the DsSs in category 1 were considered more in 1955 than 1970. 

The World Health Organization (WHO) considers the detected DsSs in this study as the common neurological disorders\footnote{\url{https://www.who.int/mental_health/neurology/neurodiso/en/}}.  This alignment between the WHO report and this study illustrates that researchers have allocated higher weight to the common DsSs in the case reports.

\section{Conclusion}
		\label{Co}
Medical experts have been contributing continuously to neurology case reports. Due to the rapid growth in the number of case reports and their overall large numbers, there is a need for a systematic analysis approach to help researchers, policymakers, and practitioners to have a large-scale understanding of the case reports. This study achieved this macro level analysis by applying text-mining and trend analysis methods on thousands neurology case reports to detect and explore high frequency DsSs and their categories from 1955 to 2017.

The results show the ability of text mining methods to investigate DsSs in a large number of medical case reports and explore them for several decades to disclose interesting patterns. Our methodology demonstrates the usefulness of computational linguistics methods to study DsSs and their trends in medical case reports. The proposed approach can be used to provide a macro level analysis of medical literature by discovering interesting patterns and tracking them in specific time frames. The results illustrated the application of text mining for detecting DsSs and their relationships, and disclosing their trends in a large number of neurology case reports. We believe that this paper proposes a systematic approach for analysis of case reports in not only neurology but also other medical fields.  

This study has some limitations. First, the collected case reports were only from the journals among the top 200 neurology journals in the Scimago Journal \& Country Rank website. In our future work, we plan to incorporate additional neurology case reports, investigate other medical research areas, and compare them. Second, we only analyzed DsSs; however, there are other entities in the case reports. Therefore, we will investigate DsSs along with chemicals, mutations, genes, and species based on time and location variables.  \\

\section{Conflict of interest}
The authors state that they have no conflict of interest.

\section{Acknowledgement}
This work is partially supported by an Advanced Support for Innovative Research Excellence (ASPIRE) grant from the Office of the Vice President for Research at the University of South Carolina. All opinions, findings, conclusions and recommendations in this paper are those of the authors and do not necessarily reflect the views of the funding agency. \\

\section*{References} 


\begin{thebibliography}{10}
	\expandafter\ifx\csname url\endcsname\relax
	\def\url#1{\texttt{#1}}\fi
	\expandafter\ifx\csname urlprefix\endcsname\relax\def\urlprefix{URL }\fi
	\expandafter\ifx\csname href\endcsname\relax
	\def\href#1#2{#2} \def\path#1{#1}\fi
	
	\bibitem{albrecht2005case}
	J.~Albrecht, A.~Meves, M.~Bigby, Case reports and case series from lancet had
	significant impact on medical literature, Journal of clinical epidemiology
	58~(12) (2005) 1227--1232.
	
	\bibitem{venes2009taber}
	D.~Venes, Taber's cyclopedic medical dictionary, FA Davis, 2009.
	
	\bibitem{nissen2014clinical}
	T.~Nissen, R.~Wynn, The clinical case report: a review of its merits and
	limitations, BMC research notes 7~(1) (2014) 264.
	
	\bibitem{sudhakaran2014role}
	S.~Sudhakaran, S.~Surani, The role of case reports in clinical and scientific
	literature, Austin J Clin Case Rep 1~(2) (2014) 1--2.
	
	\bibitem{mason2001case}
	R.~Mason, The case report--an endangered species?, Anaesthesia 56~(2) (2001)
	99--102.
	
	\bibitem{albrecht2009survey}
	J.~Albrecht, A.~Meves, M.~Bigby, A survey of case reports and case series of
	therapeutic interventions in the archives of dermatology, International
	journal of dermatology 48~(6) (2009) 592--597.
	
	\bibitem{oliveira2006critical}
	G.~J. Oliveira, C.~R. Leles, Critical appraisal and positive outcome bias in
	case reports published in brazilian dental journals, Journal of Dental
	Education 70~(8) (2006) 869--874.
	
	\bibitem{chakra2010case}
	C.~N.~A. Chakra, A.~Pariente, M.~Pinet, L.~Nkeng, N.~Moore, Y.~Moride, Case
	series in drug safety, Drug safety 33~(12) (2010) 1081--1088.
	
	\bibitem{dalmau2008anti}
	J.~Dalmau, A.~J. Gleichman, E.~G. Hughes, J.~E. Rossi, X.~Peng, M.~Lai, S.~K.
	Dessain, M.~R. Rosenfeld, R.~Balice-Gordon, D.~R. Lynch, Anti-nmda-receptor
	encephalitis: case series and analysis of the effects of antibodies, The
	Lancet Neurology 7~(12) (2008) 1091--1098.
	
	\bibitem{hoftberger2013encephalitis}
	R.~H{\"o}ftberger, M.~J. Titulaer, L.~Sabater, B.~Dome, A.~R{\'o}zs{\'a}s,
	B.~Hegedus, M.~A. Hoda, V.~Laszlo, H.~J. Ankersmit, L.~Harms, et~al.,
	Encephalitis and gabab receptor antibodies: novel findings in a new case
	series of 20 patients, Neurology 81~(17) (2013) 1500--1506.
	
	\bibitem{gallagher2007pathological}
	D.~A. Gallagher, S.~S. O'sullivan, A.~H. Evans, A.~J. Lees, A.~Schrag,
	Pathological gambling in parkinson's disease: risk factors and differences
	from dopamine dysregulation. an analysis of published case series, Movement
	disorders: official journal of the Movement Disorder Society 22~(12) (2007)
	1757--1763.
	
	\bibitem{lim1999clinico}
	A.~Lim, D.~Tsuang, W.~Kukull, D.~Nochlin, J.~Leverenz, W.~McCormick, J.~Bowen,
	L.~Teri, J.~Thompson, E.~R. Peskind, et~al., Clinico-neuropathological
	correlation of alzheimer's disease in a community-based case series, Journal
	of the American Geriatrics Society 47~(5) (1999) 564--569.
	
	\bibitem{shevell2003etiologic}
	M.~I. Shevell, A.~Majnemer, I.~Morin, Etiologic yield of cerebral palsy: a
	contemporary case series, Pediatric neurology 28~(5) (2003) 352--359.
	
	\bibitem{huang2005intracranial}
	J.~Huang, M.~J. McGirt, P.~Gailloud, R.~J. Tamargo, Intracranial aneurysms in
	the pediatric population: case series and literature review, Surgical
	neurology 63~(5) (2005) 424--432.
	
	\bibitem{mcshane1998feasibility}
	R.~McShane, K.~Gedling, B.~Kenward, R.~Kenward, T.~Hope, R.~Jacoby, The
	feasibility of electronic tracking devices in dementia: a telephone survey
	and case series, International journal of geriatric psychiatry 13~(8) (1998)
	556--563.
	
	\bibitem{wei2013pubtator}
	C.-H. Wei, H.-Y. Kao, Z.~Lu, Pubtator: a web-based text mining tool for
	assisting biocuration, Nucleic acids research 41~(W1) (2013) W518--W522.
	
	\bibitem{zhu2018understanding}
	Y.~Zhu, M.-H. Kim, S.~Banerjee, J.~Deferio, G.~S. Alexopoulos, J.~Pathak,
	Understanding the research landscape of major depressive disorder via
	literature mining: an entity-level analysis of pubmed data from 1948 to 2017,
	JAMIA Open 1~(1) (2018) 115--121.
	
	\bibitem{van2016understanding}
	A.~J. Van~Altena, P.~D. Moerland, A.~H. Zwinderman, S.~D. Olabarriaga,
	Understanding big data themes from scientific biomedical literature through
	topic modeling, Journal of Big Data 3~(1) (2016) 23.
	
	\bibitem{heimerl2014word}
	F.~Heimerl, S.~Lohmann, S.~Lange, T.~Ertl, Word cloud explorer: Text analytics
	based on word clouds, in: 2014 47th Hawaii International Conference on System
	Sciences, IEEE, 2014, pp. 1833--1842.
	
	\bibitem{sugimoto2011shifting}
	C.~R. Sugimoto, D.~Li, T.~G. Russell, S.~C. Finlay, Y.~Ding, The shifting sands
	of disciplinary development: Analyzing north american library and information
	science dissertations using latent dirichlet allocation, Journal of the
	American Society for Information Science and Technology 62~(1) (2011)
	185--204.
	
	\bibitem{leydesdorff2017co}
	L.~Leydesdorff, A.~Nerghes, Co-word maps and topic modeling: A comparison using
	small and medium-sized corpora (n< 1,000), Journal of the Association for
	Information Science and Technology 68~(4) (2017) 1024--1035.
	
	\bibitem{karami2015fuzzyiconf}
	A.~Karami, A.~Gangopadhyay, B.~Zhou, H.~Kharrazi, A fuzzy approach model for
	uncovering hidden latent semantic structure in medical text collections,
	Proceedings of the iConference. Irvine, CA.
	
	\bibitem{karami2015flatm}
	A.~Karami, A.~Gangopadhyay, B.~Zhou, H.~Kharrazi, Flatm: A fuzzy logic approach
	topic model for medical documents, in: 2015 Annual Meeting of the North
	American Fuzzy Information Processing Society (NAFIPS), IEEE, 2015.
	
	\bibitem{karami2015fuzzy}
	A.~Karami, Fuzzy topic modeling for medical corpora, Ph.D. thesis, University
	of Maryland, Baltimore County (2015).
	
	\bibitem{lu2012measuring}
	K.~Lu, D.~Wolfram, Measuring author research relatedness: A comparison of
	word-based, topic-based, and author cocitation approaches, Journal of the
	American Society for Information Science and Technology 63~(10) (2012)
	1973--1986.
	
	\bibitem{blei2003latent}
	D.~M. Blei, A.~Y. Ng, M.~I. Jordan, Latent dirichlet allocation, Journal of
	machine Learning research 3~(Jan) (2003) 993--1022.
	
	\bibitem{karami2018fuzzy}
	A.~Karami, A.~Gangopadhyay, B.~Zhou, H.~Kharrazi, Fuzzy approach topic
	discovery in health and medical corpora, International Journal of Fuzzy
	Systems 20~(4) (2018) 1334--1345.
	
	\bibitem{karami2019exploratory}
	A.~Karami, G.~Shaw, An exploratory study of (\#) exercise in the twittersphere,
	iConference 2019 Proceedings.
	
	\bibitem{karami2018characterizing}
	A.~Karami, A.~A. Dahl, G.~Turner-McGrievy, H.~Kharrazi, G.~Shaw~Jr,
	Characterizing diabetes, diet, exercise, and obesity comments on twitter,
	International Journal of Information Management 38~(1) (2018) 1--6.
	
	\bibitem{shaw2017computational}
	G.~Shaw~Jr, A.~Karami, Computational content analysis of negative tweets for
	obesity, diet, diabetes, and exercise, Proceedings of the Association for
	Information Science and Technology 54~(1) (2017) 357--365.
	
	\bibitem{karami2018characterizingtrans}
	A.~Karami, F.~Webb, V.~L. Kitzie, Characterizing transgender health issues in
	twitter, Proceedings of the Association for Information Science and
	Technology 55~(1) (2018) 207--215.
	
	\bibitem{webb2018Characterizing}
	W.~Frank, A.~Karami, K.~Vanessa, Characterizing diseases and disorders in gay
	users' tweets, in: Proceedings of the Southern Association for Information
	Systems (SAIS). Atlanta, Georgia., 2018.
	
	\bibitem{collins2018social}
	M.~Collins, A.~Karami, Social media analysis for organizations: Us northeastern
	public and state libraries case study, in: Proceedings of the Southern
	Association for Information Systems (SAIS). Atlanta, Georgia., 2018.
	
	\bibitem{karami2018us}
	A.~Karami, M.~Collins, What do the us west coast public libraries post on
	twitter?, Proceedings of the Association for Information Science and
	Technology 55~(1) (2018) 216--225.
	
	\bibitem{karami2018computational}
	A.~Karami, N.~M. Pendergraft, Computational analysis of insurance complaints:
	Geico case study, International Conference on Social Computing,
	Behavioral-Cultural Modeling, \& Prediction and Behavior Representation in
	Modeling and Simulation. Washington DC.
	
	\bibitem{karami2014exploiting}
	A.~Karami, L.~Zhou, Exploiting latent content based features for the detection
	of static sms spams, Proceedings of the 77th annual meeting of the
	association for information science and technology (ASIST). Seattle, WA.
	
	\bibitem{karami2019twitter}
	A.~Karami, V.~Shah, R.~Vaezi, A.~Bansal, Twitter speaks: A case of national
	disaster situational awareness, Journal of Information Science (2019)
	0165551519828620.
	
	\bibitem{karami2019political}
	A.~Karami, A.~Elkouri, Political popularity analysis in social media, in:
	International Conference on Information, Springer, 2019, pp. 456--465.
	
	\bibitem{karami2018mining}
	A.~Karami, L.~S. Bennett, X.~He, Mining public opinion about economic issues:
	Twitter and the us presidential election, International Journal of Strategic
	Decision Sciences (IJSDS) 9~(1) (2018) 18--28.
	
	\bibitem{chen2017revealing}
	Q.~Chen, N.~Ai, J.~Liao, X.~Shao, Y.~Liu, X.~Fan, Revealing topics and their
	evolution in biomedical literature using bio-dtm: a case study of ginseng,
	Chinese Medicine 12~(1) (2017) 27.
	
	\bibitem{wang2016text}
	S.-H. Wang, Y.~Ding, W.~Zhao, Y.-H. Huang, R.~Perkins, W.~Zou, J.~J. Chen, Text
	mining for identifying topics in the literatures about adolescent substance
	use and depression, BMC public health 16~(1) (2016) 279.
	
	\bibitem{blei2012probabilistic}
	D.~M. BLEI, Probabilistic topic models, Communications of the ACM 55~(4) (2012)
	77--84.
	
	\bibitem{zipf1949human}
	G.~K. Zipf, Human behavior and the principle of least effort.
	
	\bibitem{cao2009density}
	J.~Cao, T.~Xia, J.~Li, Y.~Zhang, S.~Tang, A density-based method for adaptive
	lda model selection, Neurocomputing 72~(7-9) (2009) 1775--1781.
	
	\bibitem{mallet}
	A.~K. McCallum, {MALLET: A Machine Learning for Language Toolkit.},
	\url{http://mallet.cs.umass.edu/topics.php} (2002).
	
	\bibitem{schlake1991symptomatic}
	H.-P. Schlake, K.-H. Grotemeyer, I.~Husstedt, G.~Schuierer, G.~Brune,
	Òsymptomatic migraineÓ: Intracranial lesions mimicking migrainous headache-a
	report of three cases, Headache: The Journal of Head and Face Pain 31~(10)
	(1991) 661--665.
	
	\bibitem{sanefuji2006moyamoya}
	M.~Sanefuji, S.~Ohga, R.~Kira, T.~Yoshiura, Moyamoya syndrome in a
	splenectomized patient with $\beta$-thalassemia intermedia, Journal of child
	neurology 21~(1) (2006) 75--77.
	
	\bibitem{iwasaki2013autopsied}
	Y.~Iwasaki, K.~Mori, M.~Ito, S.~Tatsumi, M.~Mimuro, M.~Yoshida, An autopsied
	case of progressive supranuclear palsy presenting with cerebellar ataxia and
	severe cerebellar involvement, Neuropathology 33~(5) (2013) 561--567.
	
	\bibitem{akuzawa2014secondary}
	N.~Akuzawa, T.~Osawa, M.~Totsuka, T.~Hatori, K.~Imai, Y.~Kitahara,
	M.~Kurabayashi, Secondary brain abscess following simple renal cyst
	infection: a case report, BMC neurology 14~(1) (2014) 130.
	
	\bibitem{haq2010case}
	I.~U. Haq, K.~D. Foote, W.~K. Goodman, N.~Ricciuti, H.~Ward, A.~Sudhyadhom,
	C.~E. Jacobson, M.~S. Siddiqui, M.~S. Okun, A case of mania following deep
	brain stimulation for obsessive compulsive disorder, Stereotactic and
	functional neurosurgery 88~(5) (2010) 322--328.
	
	\bibitem{mathis2013intracranial}
	D.~A. Mathis, E.~J. Stehel, J.~E. Beshay, B.~E. Mickey, A.~L. Folpe,
	J.~Raisanen, Intracranial phosphaturic mesenchymal tumors: report of 2 cases,
	Journal of neurosurgery 118~(4) (2013) 903--907.
	
	\bibitem{schuitevoerder1997proximal}
	K.~Schuitevoerder, T.~Ansved, G.~Solders, K.~Borg, Proximal myotonic myopathy,
	Acta neurologica scandinavica 96~(4) (1997) 266--270.
	
	\bibitem{macdonald1988microcystic}
	R.~L. Macdonald, J.~M. Findlay, C.~H. Tator, Microcystic spinal cord
	degeneration causing posttraumatic myelopathy: report of two cases, Journal
	of neurosurgery 68~(3) (1988) 466--471.
	
	\bibitem{tumialan2008intracranial}
	L.~Tumial{\'a}n, Y.~Zhang, C.~Cawley, J.~Dion, F.~Tong, D.~Barrow, Intracranial
	hemorrhage associated with stent-assisted coil embolization of cerebral
	aneurysms: a cautionary report., Journal of neurosurgery 108~(6) (2008) 1122.
	
	\bibitem{marras2011deep}
	C.~E. Marras, M.~Rizzi, F.~Villani, G.~Messina, F.~Deleo, R.~Cordella,
	A.~Franzini, Deep brain stimulation for the treatment of drug-refractory
	epilepsy in a patient with a hypothalamic hamartoma: case report,
	Neurosurgical focus 30~(2) (2011) E4.
	
\end{thebibliography}

	
		
		
		
\newpage

\appendix
\section{List of Journals} 
\label{appendix}

\begin{table}[htp]
	\begin{center}
		\scalebox{0.7}{
			\begin{tabular}{ll} \hline 
				Acta neurologica Scandinavica                                                                              & Journal of neurosurgery                                                     \\
				Acta neuropathologica                                                                                      & Journal of neurosurgery. Pediatrics                                         \\ 
				Annals of neurology                                                                                        & Journal of neurosurgery. Spine                                              \\ 
				Behavioural neurology                                                                                      & Journal of neurosurgical anesthesiology                                     \\ 
				BMC neurology                                                                                              & Journal of neurovirology                                                    \\ 
				Brain: a journal of neurology                                                                              & Journal of the neurological sciences                                        \\ 
				Brain injury                                                                                               & Metabolic brain disease                                                     \\ 
				Brain research                                                                                             & Multiple sclerosis and related disorders                                    \\ 
				Brain stimulation                                                                                          & Neurocritical care                                                          \\ 
				Case reports in neurology                                                                                  & Neurologic clinics                                                          \\ 
				Cephalalgia : an international journal of headache                                                         & Neurology                                                                   \\ 
				Cerebellum                                                                                                 & Neuromodulation : journal of the International Neuromodulation Society      \\ 
				Clinical neurology and neurosurgery                                                                        & Neuropathology : official journal of the Japanese Society of Neuropathology \\ 
				Clinical neuropharmacology                                                                                 & Neuropathology and applied neurobiology                                     \\ 
				CNS spectrums                                                                                              & Neurophysiologie clinique = Clinical neurophysiology                        \\ 
				Developmental medicine and child neurology                                                                 & Neuroradiology                                                              \\ 
				Epilepsia                                                                                                  & NeuroRehabilitation                                                         \\ 
				Epilepsy research                                                                                          & Neurorehabilitation and neural repair                                       \\ 
				European neurology                                                                                         & Neurosurgery                                                                \\ 
				Headache                                                                                                   & Neurosurgery clinics of North America                                       \\ 
				JAMA neurology                                                                                             & Neurosurgical focus                                                         \\ 
				Journal of child neurology                                                                                 & Neurosurgical review                                                        \\ 
				Journal of clinical and experimental neuropsychology                                                       & Pediatric neurology                                                         \\ 
				Journal of clinical neurophysiology  & Psychiatry and clinical neurosciences                                       \\ 
				Journal of clinical neuroscience           & Seminars in neurology                                                       \\ 
				Journal of geriatric psychiatry and neurology                                                              & Seminars in pediatric neurology                                             \\ 
				Journal of neuro-oncology                                                                                  & Spinal cord                                                                 \\ 
				Journal of neuro-ophthalmology     & Stereotactic and functional neurosurgery                                    \\ 
				Journal of neuroimmunology                                                                                 & The journal of spinal cord medicine                                         \\ 
				Journal of neurointerventional surgery                                                                     & The Lancet. Neurology                                                       \\ 
				Journal of neurology                                                                                       & Topics in stroke rehabilitation                                             \\ 
				Journal of neurology, neurosurgery, and psychiatry                                                         & World neurosurgery                                        \\ \hline                 
			\end{tabular}
		}
	\end{center}	
\end{table}

\end{document}